\documentclass{edm_article}
\usepackage{enumitem}
\usepackage{hyperref}
\usepackage{cleveref}
\usepackage[sort&compress,numbers]{natbib}
\begin{document}

\title{Simulating Learners' Task-Selection Strategies and System Constraints in Mastery Learning}

\numberofauthors{5}
\author{
\alignauthor Haley Noh\\
       \affaddr{New York University}\\
       \email{hhn2020@nyu.edu}
\alignauthor Aarna Chowdhary\\
       \affaddr{Carnegie Mellon University}\\
       \email{achowdha@andrew.cmu.edu}
\alignauthor Jeroen Ooge \\
       \affaddr{Utrecht University}\\
       \email{j.ooge@uu.nl}
\and
\alignauthor Vincent Aleven \\
        \affaddr{Carnegie Mellon University}\\
       \email{aleven@cs.cmu.edu}
\alignauthor Conrad Borchers \\
       \affaddr{Carnegie Mellon University}\\
       \email{cborcher@cs.cmu.edu}
}

\maketitle

\begin{abstract}
 Intelligent Tutoring Systems often grant learners shared control over skill and problem selection. This choice brings motivational and metacognitive benefits. At the same time, past literature suggests that learners exhibit diverse preferences and strategies in selecting tasks, for instance, by avoiding challenge. Although underexplored, differences in learner task-selection strategies may interact with mastery learning systems that optimize task-selection based on estimated knowledge, potentially leading to undesirable student-level differences in learning outcomes. Algorithmic constraints on problem selection may help mitigate this issue. However, this possibility has not been comprehensively explored in prior work, in part because testing such constraints in real-world classrooms is costly. We propose a simulation-based framework to observe how varying learner task-selection strategies combined with system constraints shape mastery learning efficiency. Using interaction data from 261 students across two mathematical domains with different problem structures (equation solving, graph interpretation), we simulate common task-selection strategies such as Weakness Targeting and Interleaving, grounded in prior literature. We then evaluate how these strategies affect overpractice as a common measure of mastery learning efficiency. Results show substantial variability in efficiency across strategies, with risk-averse strategies producing higher levels of overpractice, especially for more complex multi-step problems. Targeted system constraints significantly reduce these inefficiencies for maladaptive strategies while having minimal impact on already efficient strategies. Together, these findings demonstrate how simulation grounded in real student data can support data-driven redesign of shared-control tutoring systems by identifying when and where constraints are most beneficial prior to classroom deployment.
\end{abstract}


\section{Introduction}
Mastery-based learning is widely adopted in Intelligent Tutoring Systems (ITS) to personalize instruction and ensure learners demonstrate proficiency in prerequisite skills before advancing \cite{Pelanek2017, Koedinger2015DataMining}. Although ITS typically do not involve learners in the adaptive task-selection process, prior work has shown that involving learners by allowing them to express or act on task-selection preferences can yield motivational and metacognitive benefits, including increased engagement, perceived autonomy, and reflection on learning progress \cite{scheiter2007learner, Corbalan2008Adaptation, long2017, bull2020olm}. However, differences in learner task-selection strategies can introduce substantial inefficiencies in mastery-based learning, particularly when learners choose problems that involve skills they have already mastered or avoid challenging ones \cite{Pelanek2017, xia2025}. Prior research has shown that learners adopt diverse task-selection strategies such as Strength Targeting, Weakness Targeting, Interleaving, or Blocking \cite{butler2005, foster2019, long2017}. While some strategies align well with mastery-based learning objectives, others can result in excessive overpractice and delayed progression, particularly in domains with multi-step problem structures where mastered skills frequently reappear.

Studying how different learner strategies paired with task-selection constraints through classroom experiments is costly, time-consuming, and may expose learners to suboptimal learning experiences. Alternatively, simulation studies with real learner interaction data enable systematic and efficient comparison of learner strategies and instructional design choices before deployment \cite{Pelanek2017, xia2025, Koedinger2023}. By modeling how different forms of learner control and system guidance shape learning trajectories, simulation offers a scalable method for identifying inefficiencies and testing corrective mechanisms before they affect real students.

We present a simulation-based framework for analyzing how learner task-selection strategies interact with system constraints to shape mastery-based learning efficiency in ITS. Using interaction data from two mathematical domains (equation solving and graph interpretation), we simulate common learner strategies and evaluate their effects using measures of overpractice. This approach enables controlled comparison of learning control and system-control configurations, providing insights into when targeted scaffolding is necessary to mitigate inefficiencies while preserving learner choice. In sum, our research question is as follows:

\textbf{RQ.} How do different learner task-selection strategies, combined with system constraints, affect the efficiency of mastery-based learning in ITS?

\section{Related Work}

\subsection{Task-Selection Strategies in ITS}
Mastery learning frameworks advance students only after demonstrating mastery of prerequisite skills, using criteria based on performance \cite{Pelanek2017}. This ensures competence but can create inefficiencies, as some skills become overpracticed while others are neglected \cite{xia2025}. In many modern ITS, learners are increasingly involved in selecting which skills or problems to practice, often supported by learning analytics and open learner models (OLMs) that make progress and mastery visible \cite{long2017, bull2020olm}. Allowing learners to make such choices is motivated by self-determination theory, which emphasizes autonomy as a key driver of intrinsic motivation, engagement, and persistence in learning \cite{urhahne2023}. Prior work in educational data mining and ITS shows that learners adopt diverse task-selection strategies \cite{butler2005}. Studies show that when learners freely select skills or problems, they pursue strategies of uneven effectiveness, indicating that control alone rarely improves outcomes and can even hinder them if left unsupported \cite{scheiter2007learner}. Some choose familiar tasks, others target weaknesses \cite{butler2005}; some interleave across skills, while others block practice within a single skill, often shaped by confidence and goals \cite{foster2019}. These strategies can nurture persistence and reflection, but also cause inefficiencies, such as overpractice of mastered skills or avoidance of challenges. Understanding how learners naturally manage practice is therefore essential, both to amplify productive strategies and to counter maladaptive ones. Despite this, little research has systematically examined how such strategies interact with system-level supports like constraints, guidance, or explanations.

\subsection{System Constraints and Simulation}
To address variability in learner behavior, ITS increasingly combines learner choice with system-level constraints that guide task-selection while preserving agency, a paradigm often described as shared student-system control over task-selection. Prior work shows that constrained problem-selection and guided choice can reduce overpractice and improve efficiency without fully removing learner control \cite{xia2025, yan2024control, long2017}. However, evaluating how different learner strategies interact with such constraints through classroom studies is costly and risks exposing students to suboptimal learning conditions. 

Simulation studies have therefore become increasingly valuable in education research, especially where in vivo experimentation is infeasible or cost-prohibitive \cite{maclellan2025modelhumanlearnerscomputational}. In addition, deploying poor algorithms can be harmful to participants. Simulation enables systematic exploration of both learner strategies and shared-control mechanisms, such as weightings, recommendations, and explanation styles, in a controlled environment. Prior research demonstrates that simulated learners can reproduce realistic learning trajectories and support the development of novel optimization techniques for tutoring systems \cite{xia2025}. Rather than aiming to predict individual learning outcomes, simulation enables comparative analysis of alternative design configurations under consistent assumptions \cite{Pelanek2017, Koedinger2023}. In this work, we build on this line of research by using simulation to examine when shared-control constraints are necessary to mitigate inefficiencies arising from specific learner task-selection strategies.

\section{Methods and Materials}
\subsection{Datasets}
We used two datasets to specify model parameters and enhance validity. Following \cite{xia2025}, we defined the most common solution pathways in the data and based the simulations on them \cite{kaeser2024}. The first dataset is on equation-solving and used an open-source dataset available on PSLC DataShop (datasets \#5549 and \#5604), collected from IRB-approved middle school studies using the APTA ITS for linear equation solving \cite{koedinger2010data, borchers2024combining}. The dataset includes students in grades 6--8 attending two public schools in the eastern United States and contains over 10,000 step-level interactions with multi-step equation problems \cite{Koedinger2015DataMining}. To model authentic first-attempt problem solving, only initial responses were retained, and steps involving hints were treated as incorrect \cite{Koedinger2015DataMining}. The step-level structure and repeated use of skills enable reconstruction of typical solution paths, which form the basis for simulating learning trajectories and task-selection strategies. This dataset contains 58 problems and 10 skills. The average number of skills that appeared in a problem was $5.67 \pm 2.47$.

We additionally used another DataShop dataset based on graph interpretation (\#5360) \cite{koedinger2010data}. This dataset consisted of classroom transaction data from the Mathtutor ITS, collected through IRB-approved studies with 97 ninth-grade students in a U.S. high school. Students alternated between paper practice and Mathtutor across three units on linear graphs, with the problem decomposed into step-level interactions (though we only used data from the ITS condition). The dataset represents another domain of multi-step problem solving where students may engage in overpractice to improve generalizability beyond the equation-solving data. This dataset contains 31 problems and 13 skills. The average number of skills that appeared in a problem was $8.57 \pm 8.41$.

The datasets differ in problem length and problem-skill composition: graph interpretation involves more skills and steps per problem than equation solving. Because these factors affect overpractice \cite{xia2025}, this enables evaluation of skill and problem-selection strategies across varied problem types.
    
\subsection{Learner Simulation Methods}
We use a simulation-based approach to model student learning under different task-selection strategies in a mastery-based ITS. We specifically use two widely used student modeling approaches in educational data mining research. We use the Additive Factors Model (AFM; \cite{cen2006afm}) to simulate step-level performance and Bayesian Knowledge Tracing (BKT; \cite{corbett1994bkt}) to model knowledge evolution over time. The simulation framework is built on real student interaction data using open-source learner simulation libraries described in \citet{xia2025}.

To estimate skill difficulty and learning parameters for our two datasets, we fit both AFM and BKT to the PSLC DataShop and MathTutor ITS data collections and use the learned parameters to drive the student simulations. Each problem is associated with a typical solution path derived from the most frequent skill-application sequences in the data \cite{Stamper2013}. During simulation, student responses are sampled at each step according to the AFM-predicted probability of correctness, and both AFM and BKT states are updated after each interaction. The resulting BKT mastery estimates are then used to guide subsequent problem-selection. Our simulation and analysis code is open-source.\footnote{\url{https://github.com/conradborchers/sim-skill-selection}}.

AFM models the probability of a correct response as a logistic function of a global intercept, a student ability term, and additive skill-specific difficulty and practice effects. Specifically, calculations are made using information such as the set of skills associated with a problem step, the difficulty of these skills, and the attempts the student has previously made for these skills. BKT models student learning as a hidden Markov process in which each skill has a binary latent state: either learned or unlearned.  The model updates mastery using observed responses and four parameters: Prior knowledge estimates ($p_\text{init}$), learning rate ($p_\text{learn}$), guess rate ($p_\text{guess}$), and slip rate ($p_\text{slip}$).

Upon receiving a student response, BKT updates the estimated mastery level using Bayes' rule. A correct answer increases the mastery probability, which adjusts the potential for guessing, while an incorrect answer decreases the probability, accounting for possible slips. A skill is considered mastered once the posterior probability surpasses a predefined mastery threshold, in our case 95\%, consistent with prior ITS research \cite{borchers2025}. The initial values of the BKT parameters are based on the following default values in TutorShop: $p_\text{init} =0.25$, $p_\text{learn} = 0.22$, $p_\text{guess} = 0.2$, and $p_\text{slip} = 0.1$.

\subsection{Simulation Overview}
 Each simulation allows 1,000 learners to progress through a set of skills until mastery is achieved. The simulation is designed to support controlled comparison of learner strategies and system constraints under standardized conditions. 
The simulation models mastery learning as a repeated problem-level decision cycle. The iterative cycle of task-selection, problem-solving, and mastery updating defines a learning path for each simulated learner. Specifically, a learning path is represented as the sequence of skills selected and problems attempted over time in combination with the changing mastery estimates. As task-selection decisions are made based on the learner's current knowledge state and updated after each simulated problem attempt, the resulting trajectories reflect the learner's emergent path through an adaptive task sequence. Between problems, a simulated learner selects a skill to practice based on a predefined task-selection strategy, optionally shaped by system-imposed constraints on the available skill choices. Once a skill is selected, the system chooses a problem that uses that skill, potentially applying additional problem-selection constraints. This modeling choice reflects a common shared-control setup in mastery-based ITS, where learners select skills, and the system selects corresponding problems, but does not capture all possible forms of task-selection \cite{huang2020designloop,borchers2025}.

The simulated learner then attempts the selected problem, which may consist of multiple steps, each mapping to at least one skill. The steps are simulated sequentially, with performance generated probabilistically based on the learner's current knowledge state and step difficulty. Mastery estimates are updated after each step, but the learner does not make additional choices until the entire problem has been completed. After the problem attempt concludes and mastery estimates are updated, the learner again selects a skill for the next problem based on their updated skill mastery estimates, and the cycle repeats.

\begin{figure}
    \centering
    \includegraphics[width=1 \linewidth]{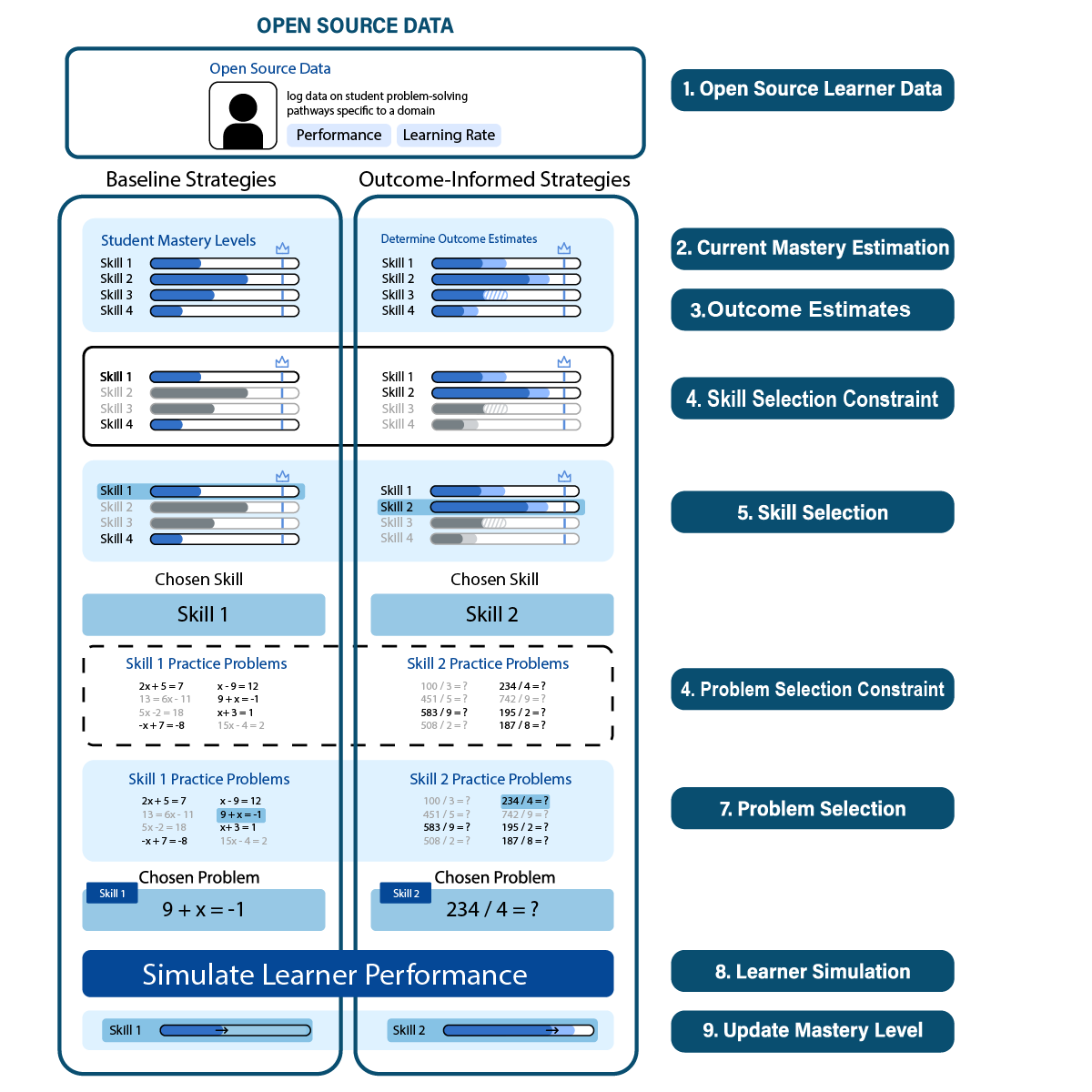}
    \caption{Simulation Methodology Summary.}
    \Description{Flow diagram comparing two learner-simulation strategies using open-source learner data. 
The top box shows log data on student problem-solving pathways, including performance and learning-rate measures. 
Two side-by-side pipelines are shown: a baseline strategy and an outcome-informed strategy. 
Both estimate current student mastery across four skills, apply skill-selection constraints, choose a skill, apply problem-selection constraints, choose a practice problem, simulate learner performance, and update mastery. 
The baseline pipeline selects Skill 1 and the problem ``9 + x = -1,'' while the outcome-informed pipeline uses outcome estimates to select Skill 2 and the problem ``234 / 4 = ?''. 
A numbered list on the right labels the process stages from open-source learner data through mastery update.}
    \label{fig:design}
\end{figure}

This structure mirrors the way that learners can select tasks when interacting with shared-control mastery-based ITS, where strategic decisions are made between problems rather than within them \cite{borchers2025,long2017}. By grounding step-level performance and mastery updates in parameters estimated from real student data, the simulation produces realistic learning trajectories while enabling controlled comparison of how different learner task-selection strategies and system constraints affect mastery learning efficiency. Figure~\ref{fig:design} summarizes this iterative problem-selection-practice loop.

Below, we elaborate on the individual components of the simulation framework:
\begin{enumerate}[leftmargin=*]
    \item \textbf{Open Source Learner Data} We begin by using open-source log data from both middle school and high school math ITS, which contains detailed records of student interactions and performance across a range of equation-solving and graph-interpretation skills, respectively (see Section \ref{fig:design}). 
    \item \textbf{Current Mastery Estimation} The estimated mastery levels for each skill present in the dataset are calculated after every practice opportunity.  
    \item \textbf{Outcome Estimates} For strategies that specifically rely on outcome-estimations, we compute the projected mastery outcomes under three scenarios: best case, usual case, and worst case.
    \item \textbf{Task-Selection Guidance} Front-end task-selection guidance simulates UI cues by restricting the pool of selectable skills to those closer or further from the mastery threshold, shaping learner decisions without altering algorithms.
    \item \textbf{Task-Selection}  A skill is selected using either a general strategy (e.g., Strength Targeting, Interleaving) or an Outcome-informed strategy (e.g., Maximize Usual Case, Minimize Worst Case Loss), simulating different learner decision-making behaviors.
    \item \textbf{Problem-Selection Constraint} Once a skill is selected, a system control constraint introduces bias in the subsequent problem-selection process based on problem difficulty, modeling how backend design influences the learning path.
    \item \textbf{Problem-Selection} A specific problem is randomly drawn from the available pool associated with the selected skill, either randomly or following any imposed system control constraints. 
    \item \textbf{Learner Simulation} The simulated student attempts the selected problem step, and performance is determined probabilistically using AFM, based on current skill mastery and step difficulty.
    \item \textbf{Update Mastery Level} Following the simulated attempt, mastery estimates are updated using BKT, and the simulation loop repeats until all skills reach mastery or a stopping condition is met.

\end{enumerate}

\subsection{Task-Selection Strategies}
We model learner task-selection using predefined decision rules that determine which skill a simulated learner chooses to practice at each step. The eight selected strategies are designed to reflect commonly observed patterns of learner behavior in mastery-based tutoring systems and are applied consistently across all simulation conditions.

\subsubsection{Baseline Task-Selection Strategies}
We implement four baseline strategies drawn from prior work on learner practice behavior. \textbf{Strength Targeting} prioritizes skills that are close to mastery, while \textbf{Weakness Targeting} prioritizes skills with the lowest estimated mastery. \textbf{Interleaving} alternates practice across multiple skills, whereas \textbf{Blocking} concentrates practice on a single skill until mastery is achieved. These strategies capture a range of common learner approaches to managing practice and have been widely studied in educational psychology and ITS research. Finally, a \textbf{Random} strategy selects skills at random and serves as a control condition.

\subsubsection{Outcome-Informed Strategies}
In addition to baseline strategies, we model decision rules that select skills based on projected mastery outcomes under different assumptions about performance. \textbf{Maximize Usual Case Improvement} selects the skill with the greatest expected mastery gain, while \textbf{Maximize Usual Case Outcome} selects the skill with the highest projected post-practice mastery. \textbf{Minimize Worst Case Loss} represents a risk-averse strategy that selects the skill with the smallest potential decrease in mastery under unfavorable performance. This strategy is risk-averse because it prioritizes minimizing potential negative mastery outcomes over maximizing expected gains.

Together, these strategies enable systematic comparison of how different decision-making behaviors interact with system-level constraints to influence learning efficiency in mastery-based tutoring systems.

\subsection{System Constraints}
To examine how system design interacts with task-selection strategies, we introduce two system constraints that shape learning while preserving learner control over skill prioritization.

\subsubsection{Task-Selection Constraints}
Task-selection constraints limit the set of skills available for selection at each decision point based on current mastery estimates. We implement two variants. The \textit{closer-to-mastery} constraint restricts selection to skills with proficiency levels nearest to, but below, the mastery threshold. The \textit{further-from-mastery} constraint restricts selection to skills with lower proficiency levels, representing a weaker understanding. These constraints reduce learners' choice set without prescribing a specific selection strategy, allowing learners to retain control within a bounded decision space. We also simulate learners' interactions under unconstrained task-selection to provide a baseline for comparison. 

\subsubsection{Problem-Selection Constraints}
 Problem-selection constraints influence which problem is delivered after learners select a skill. Rather than selecting problems uniformly at random, the system applies biased sampling within the chosen skill. Problem difficulties are determined using a weighted average of the learners' past performance with skills present in problem steps. Under the \textit{prefer-easier constraint}, problems associated with lower difficulty skills are assigned higher weight, while under the \textit{prefer-harder constraint}, problems associated with higher difficulty skills are given higher weight. Once all the problem difficulty scores are assigned, they are normalized to form a multinomial distribution from which weighted random sampling is used to select a problem. These constraints operate independently of task-selection and affect only the execution of practice within a chosen skill.

\subsection{Metric: Overpractice}
We evaluate the impact of system constraints in ITS using \textbf{overpractice} as the primary outcome. Overpractice refers to continued practice of skills that have already reached mastery, a phenomenon that commonly arises in mastery-based tutoring systems \cite{Pelanek2017, xia2025}. This particularly occurs when problems involve multiple skills or steps, as problems primarily based on already proficient skills may be chosen for including an unmastered component \cite{Koedinger2015DataMining}. Overpractice has been studied as an inefficiency in mastery-based ITS, as it increases time on task without proportional learning benefits and can slow progression through the learning process \cite{Pelanek2017, xia2025, Koedinger2023}. For each strategy-constraint condition, simulation runs produced data on the overpractice aggregated across all skills present in the dataset, yielding descriptive statistics, including mean and standard deviation. In addition to comparing means and effect sizes using Cohen's d, the analysis focused on determining which strategies and constraints resulted in the most efficient mastery learning, indicated by reduced overpractice. The findings helped identify optimal combinations of strategies and constraints that minimized unnecessary practice while maintaining mastery.

\section{Results}
The results presented in Figures \ref{fig:baseline} and \ref{fig:mwl} summarize the average overpractice observed under different task-selection strategies and system constraint conditions. These values are computed by aggregating overpractice across all simulated learners and skills. This allows for direct comparison of learning efficiency across strategies and datasets.  
\subsection{Performance Across Strategies}
\begin{figure}
    \centering
    \includegraphics[width=1 \linewidth]{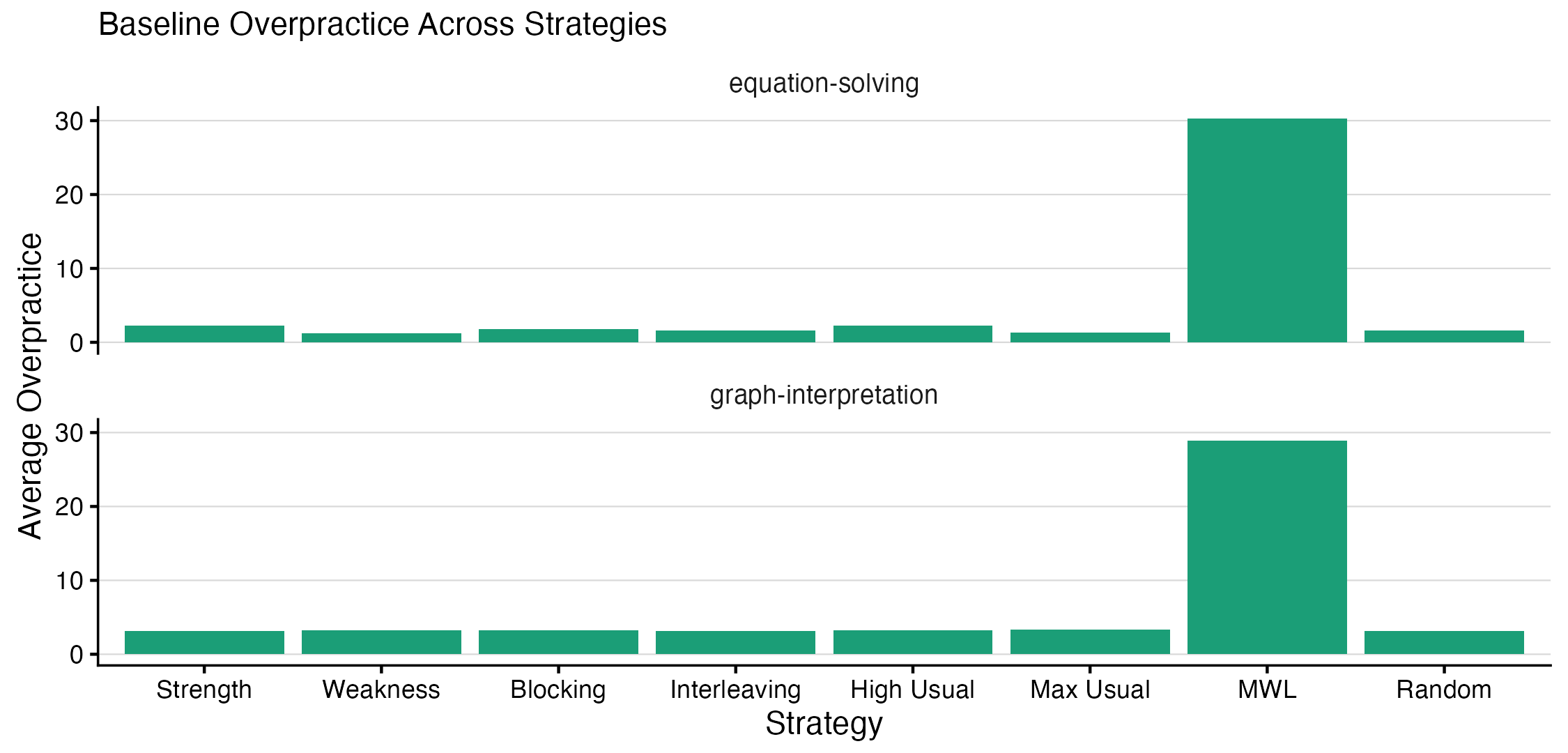}
    \caption{Baseline Average Overpractice Across Strategies for Both Datasets. MWL denotes the Minimize Worst Case Loss strategy.}
    \Description{Two bar charts compare average overpractice across instructional strategies in the domains of equation solving and graph interpretation. 
The x-axis lists eight strategies: Strength, Weakness, Blocking, Interleaving, High Usual, Max Usual, MWL, and Random. 
The y-axis shows average overpractice. 
In both domains, most strategies produce low and similar overpractice values, generally between 1 and 4. 
The MWL strategy is a strong outlier, with substantially higher overpractice—approximately 30 for equation solving and 29 for graph interpretation. 
All other strategies remain clustered near the bottom of the scale, indicating comparatively limited overpractice.}
    \label{fig:baseline}
\end{figure}
Figure \ref{fig:baseline} compares baseline overpractice levels across all task-selection strategies without system constraints. Across both datasets, minimizing worst case loss was a clear outlier in terms of its average amount of overpractice. Without system constraints, using the Minimize Worst Case Loss strategy produced overpractice levels of 30.28 times in equation-solving and 28.93 times on average in graph-interpretation (Figure \ref{fig:baseline}). In contrast, the average of the other seven strategies' overpractice amounts were $1.733 \pm 0.383$ and $3.223 \pm 0.050$ for the equation-solving and the graph-interpretation data sets, respectively. Additionally, the second-worst performing strategy for the equation-solving dataset was the Strength Targeting strategy. This indicates that informed risk-averse strategies are substantially less efficient than other strategies in self-guided systems.

On the other hand, in the equation-solving dataset, Weakness Targeting and Maximize Usual Case Improvement performed slightly better than other strategies, yielding the lowest overpractice of 1.26 and 1.29 average practice opportunities, respectively (Figure \ref{fig:baseline}). This suggests that prioritizing less-mastered skills is a comparatively efficient approach. 

 \subsection{Incorporating System Constraints}
\begin{figure}
    \centering
    \includegraphics[width=1 \linewidth]{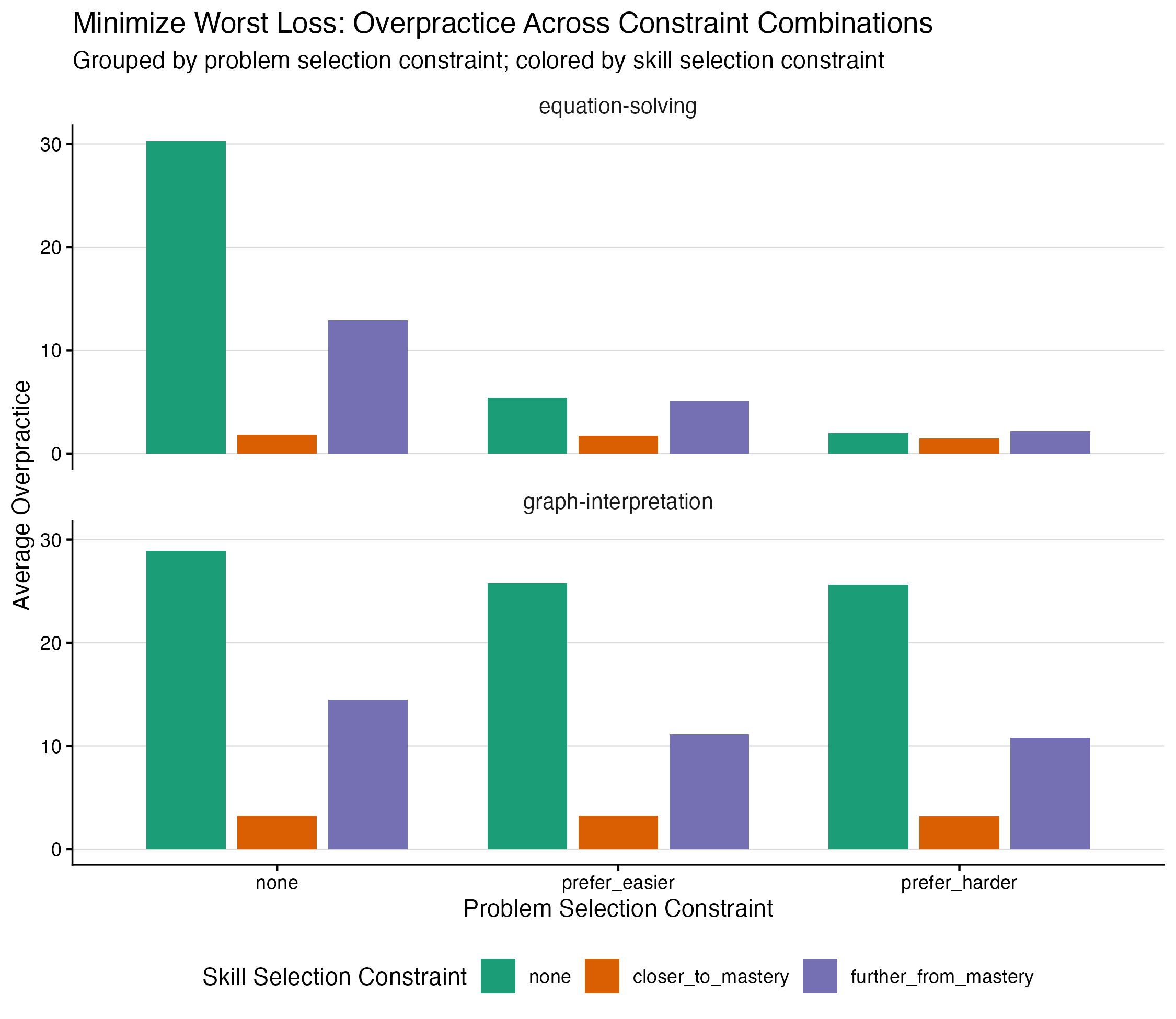}
    \caption{Average Overpractice for the Minimize Worst Case Loss Strategy with Task-Selection and Problem-Selection Constraints.}
    \label{fig:mwl}
    \Description{Two grouped bar charts show average overpractice under combinations of problem-selection and skill-selection constraints for the Minimize Worst Loss strategy. 
The top chart shows equation solving and the bottom chart shows graph interpretation. 
The x-axis groups results by problem-selection constraint: none, prefer easier, and prefer harder. 
Within each group, colored bars represent skill-selection constraints: none, closer to mastery, and further from mastery. 
The y-axis shows average overpractice.

For equation solving, the highest overpractice occurs when no constraints are applied, especially with no skill-selection constraint, at about 30. 
Adding either problem-selection or skill-selection constraints substantially reduces overpractice. 
The lowest values, around 1 to 2, occur when selecting skills closer to mastery.

For graph interpretation, overpractice remains high whenever no skill-selection constraint is used, with values around 25 to 29 regardless of the problem-selection constraint. 
Using the closer-to-mastery skill constraint reduces overpractice to about 3 across all problem-selection conditions. 
The further-from-mastery constraint produces intermediate overpractice values around 11 to 15. 
Overall, skill-selection constraints have a larger effect on reducing overpractice than problem-selection constraints.}
\end{figure}
Figure \ref{fig:mwl} illustrates how task-selection and problem-selection constraints affect overpractice for the Minimize Worst Case Loss strategy. 
Introducing task-selection constraints significantly moderated outcomes for otherwise inefficient strategies. Most notably, when the risk-averse strategy (Minimize Worst Case Loss) was guided toward skills closer to mastery, the excess overpractice was sharply reduced from 30.28 to 1.83 times in equation-solving and 28.93 to 3.22 times in graph-interpretation relative to the unconstrained baseline, with large respective Cohen's d effect sizes of 1.87 and 1.86 (Figure \ref{fig:mwl}). Task-selection constraints that prioritized skills further from mastery also improved efficiency, reducing overpractice to 12.88 and 14.46, respectively. By contrast, most other strategies were relatively unaffected by these constraints. Blocking, Interleaving, Strength Targeting, and Weakness Targeting all maintained stable levels of overpractice across conditions. Overall, the results
indicate that specific forms of system control exert their strongest effects on risk-averse learners, while already efficient strategies show minimal change.

Problem-selection constraints also had a considerable effect on reducing inefficiencies in simulated learning. When paired with the Prefer Harder constraint, Minimize Worst Case Loss showed a greatly reduced overpractice value from 30.28 to 1.98 times in the equation-solving dataset, a large Cohen's d effect size value of 1.86, leveling the other strategies without the constraint (Figure \ref{fig:mwl}). This suggests that system control constraints can help counterbalance risk-averse tendencies by increasing exposure to more challenging problems. The equation-solving dataset, where multi-step structures magnified redundancy, produced a stronger corrective effect when paired with the Prefer Harder constraint compared to graph-interpretation.

\section{Discussion}
As learner control becomes increasingly common in ITS and other adaptive learning environments \cite{scheiter2007learner, Corbalan2008Adaptation, brusilovsky2024}, understanding how differences in task-selection behavior affect learning efficiency is critical. Rather than assuming that learner choice uniformly benefits all students, our results demonstrate substantial variability in outcomes across task-selection strategies. Using a simulation derived from real student interaction data, this study shows how certain learner decisions can systematically hinder mastery learning, while others align well with learning objectives.

\subsection{Strategy Variability and Validation}
Across both datasets, weakness-targeting strategies consistently produced efficient learning paths with minimal overpractice. In contrast, the risk-averse strategy that minimizes worst case loss, which tends to prioritize skills with already high estimated mastery, emerged as a clear outlier and resulted in substantially higher levels of overpractice. 
This pattern may be explained by the fact that stronger skills are closer to mastery and involve lower expected errors and mastery loss, leading the risk-averse strategy to favor already well-practiced skills and thereby induce redundant practice.
Notably, this form of risk aversion has been documented in real learners and is therefore likely to generate meaningful differences in learning efficacy across students, depending on whether they exhibit this behavior \cite{borchers2025}. Next we discuss how constraints on task-selection can mitigate these differences.

\subsection{System Constraints Mitigate Inefficiencies}
Our findings show that system constraints on task-selection are more effective when applied selectively. Task-selection and problem-selection constraints substantially reduced inefficiencies for the risk-averse strategy, bringing its performance to levels similar to those of other, more efficient strategies overall. At the same time, these constraints had little to no effect on strategies that were already efficient, such as Weakness Targeting or Interleaving. This asymmetry suggests that applying the same constraints to all learner types is unnecessary (and that student preferences for specific skills can be honored for some students more than others without sacrificing mastery efficiency). Instead, simulation enables identification of which learner behaviors warrant targeted system constraints, allowing system designers to intervene only where inefficiencies are likely to arise. Such selective constraint application aligns with common practice in ITS \cite{Pelanek2017, huang2020designloop}, where system constraint is routinely adapted based on learner state rather than applied uniformly. An open question for future work is whether students would be able to recognize different levels of system constraints on task-selection (some being invisible to them but potentially felt during problem completion). Overly prominent system constraints may undermine student autonomy perceptions and thereby lower motivation, ultimately hindering learning \cite{borchers2025student}. However, because our constraints are designed to preserve shared control by bounding rather than removing learner choice, they are intended to mitigate inefficiencies while maintaining the motivational benefits associated with autonomy.

\subsection{Equity and Shared Control}
Prior work based on large-scale evidence has shown that students exhibit large differences in their level of engagement and persistence in various adaptive learning systems \cite{Koedinger2023, holt2024five}. Here, we turn to another, underinvestigated aspect of student outcomes in mastery learning: the impact of decision-making strategies when choosing tasks. While our findings suggest that constraining control may lead to more equitable benefits of mastery learning, it is also worth highlighting that there are benefits to student control, particularly with regards to the aforementioned persistence differences highlighted in prior work \cite{Koedinger2023, holt2024five}. In particular, past research notes motivational benefits of control related to engagement, motivation, and metacognitive development. By contrast, targeted system constraints offer a middle ground where preserving learner choice while mitigating extreme inefficiencies that disproportionately affect certain learners \cite{scheiter2007learner, long2017, Corbalan2008Adaptation}.

\subsection{Simulation to Close the Loop}
More broadly, this work contributes a novel framework and evidence of how simulations can support close-the-loop experimentation in educational data mining by serving as a bridge between data analysis and learning system redesign \cite{huang2020designloop}. Grounded in real student interaction data and using standard EDM models such as AFM and BKT, our simulation enables systematic evaluation of how alternative design choices, specifically different levels of learner control and system constraints, shape learning efficiency prior to classroom deployment. This aligns with prior work on design-loop adaptivity, which emphasizes using data-driven analyses to inform and prioritize redesign decisions that can later be validated through in vivo experimentation \cite{huang2020designloop}. We treat simulation as a complementary pre-deployment tool that reduces risk and helps identify which learner behaviors and system interventions warrant targeted classroom evaluation.

Additionally, the simulation was reproducible across datasets of student interaction data spanning different problem domains. Common anticipated trends—such as inefficiencies arising from risk-averse strategies—were consistently observed across both datasets. Notably, overpractice outcomes for equation-solving tasks exhibited greater variability than those observed in other domains. This finding aligns with the notion that equation-solving problems often admit multiple solution paths involving distinct underlying skills, in contrast to graph-interpretation tasks. Moreover, equation-solving problems frequently exhibit highly recursive structures with greater skill repetition, which may further contribute to the observed variability \cite{xia2025}.

\subsection{Limitation and Future Work}
This study models a limited set of learner task-selection strategies and system constraints grounded in prior empirical work, and does not capture the full range of motivational, metacognitive, or affective factors that may influence learner behavior. Our simulation focuses on a task-selection-first setup and does not model alternative task-selection paradigms (e.g., direct problem-selection, multi-task-selection, or fully system-driven sequencing), though the framework could be extended to support these regimes. Learner strategies are modeled as static, whereas real learners adapt their strategies over time in response to feedback or outcomes. We leave the investigation of dynamic combinations of strategies to future work.
Future work could also extend our framework by incorporating models of motivation and self-regulated learning to examine how different forms of shared control affect engagement and metacognitive processes. Finally, the simulation-based approach can inform iterative, close-the-loop system redesign by identifying promising constraint strategies prior to classroom deployment and refining models using additional learner data across domains \cite{huang2020designloop}. 

\section{Conclusion}
This study contributes a novel simulation-based educational data mining framework for examining how learner task-selection strategies and system constraints jointly shape mastery learning efficiency in ITS. Grounded in real student data and implemented using standard EDM models, our simulations reveal substantial variability in the efficiency of different learner task-selection strategies. In particular, while some strategies align with mastery learning objectives, others can lead to extreme inefficiencies in practice. Our results show that targeted system constraints can mitigate these inefficiencies without removing learner control, while having minimal impact on strategies that are already effective. This highlights the value of selectively deploying scaffolding mechanisms rather than adopting uniform control policies, and underscores the importance of accounting for learner strategy diversity in the design of adaptive learning systems. Broadly speaking, this study demonstrates how simulation can support close-the-loop data-driven redesign in educational data mining. By enabling systematic evaluation of alternative design choices prior to classroom deployment, simulation helps reduce risk, accelerate iteration, and prioritize which interventions merit further classroom testing. As learner-controlled systems become more widespread, these methods will be increasingly important for ensuring that the benefits of educational technology are distributed equitably without amplifying disparities in learning outcomes. Practically, this work suggests that ITS designers seeking to implement shared control should move toward selectively or adaptively applying constraints that target inefficient learner task-selection strategies. Our findings highlight the value of simulation as a low-cost and scalable tool for evaluating ITS design choices, enabling researchers to test and refine potentially inequitable or suboptimal interventions before deploying them in real learning environments.

\section*{Acknowledgments}
This research was funded by the Institute of Education Sciences (IES) of the U.S. Department of Education (Award \#R305A220386). H. Noh was supported by the National Science Foundation (NSF) Research Experiences for Undergraduates (REU) program. A. Chowdhary was supported by the Summer Undergraduate Research Apprenticeship (SURA) program at Carnegie Mellon University.

\bibliography{sigproc}  %

\begin{thebibliography}{24}
\providecommand{\natexlab}[1]{#1}
\providecommand{\url}[1]{\texttt{#1}}
\expandafter\ifx\csname urlstyle\endcsname\relax
  \providecommand{\doi}[1]{doi: #1}\else
  \providecommand{\doi}{doi: \begingroup \urlstyle{rm}\Url}\fi

\bibitem[Borchers et~al.(2024)Borchers, Yang, Lin, Rummel, Koedinger, and Aleven]{borchers2024combining}
Conrad Borchers, Kexin Yang, Jionghao Lin, Nikol Rummel, Kenneth~R. Koedinger, and Vincent Aleven.
\newblock Combining dialog acts and skill modeling: What chat interactions enhance learning rates during ai-supported peer tutoring?
\newblock In Benjamin Paa{\ss}en and Carrie~Demmans Epp, editors, \emph{Proceedings of the 17th International Conference on Educational Data Mining}, pages 117--130, Atlanta, Georgia, USA, July 2024. International Educational Data Mining Society.
\newblock \doi{10.5281/zenodo.12729784}.

\bibitem[Borchers et~al.(2025{\natexlab{a}})Borchers, Ooge, Peng, and Aleven]{borchers2025}
Conrad Borchers, Jeroen Ooge, Cindy Peng, and Vincent Aleven.
\newblock How learner control and explainable learning analytics about skill mastery shape student desires to finish and avoid loss in tutored practice.
\newblock In \emph{Proceedings of the 15th International Learning Analytics and Knowledge Conference}, pages 810--816, 2025{\natexlab{a}}.
\newblock \doi{10.1145/3706468.3706484}.

\bibitem[Borchers et~al.(2025{\natexlab{b}})Borchers, Peng, Lyu, Carvalho, Koedinger, and Aleven]{borchers2025student}
Conrad Borchers, Cindy Peng, Qianru Lyu, Paulo~F Carvalho, Kenneth~R Koedinger, and Vincent Aleven.
\newblock Student perceptions of adaptive goal setting recommendations: a design prototyping study.
\newblock In \emph{International Conference on Artificial Intelligence in Education}, pages 244--251. Springer, 2025{\natexlab{b}}.
\newblock \doi{10.1007/978-3-031-98462-4_31}.

\bibitem[Brusilovsky(2024)]{brusilovsky2024}
Peter Brusilovsky.
\newblock Ai in education, learner control, and human-ai collaboration.
\newblock \emph{International Journal of Artificial Intelligence in Education}, 34\penalty0 (1):\penalty0 122--135, 2024.
\newblock \doi{10.1007/s40593-023-00356-z}.

\bibitem[Bull(2020)]{bull2020olm}
Susan Bull.
\newblock There are open learner models about!
\newblock \emph{IEEE Transactions on Learning Technologies}, 13\penalty0 (2):\penalty0 425--448, 2020.
\newblock \doi{10.1109/TLT.2020.2978473}.

\bibitem[Butler et~al.(2005)Butler, Beckingham, and Lauscher]{butler2005}
Deborah~L Butler, Beverly Beckingham, and Helen J~Novak Lauscher.
\newblock Promoting strategic learning by eighth-grade students struggling in mathematics: A report of three case studies.
\newblock \emph{Learning disabilities research \& practice}, 20\penalty0 (3):\penalty0 156--174, 2005.
\newblock \doi{10.1111/j.1540-5826.2005.00130.x}.

\bibitem[Cen et~al.(2006)Cen, Koedinger, and Junker]{cen2006afm}
Hao Cen, Kenneth Koedinger, and Brian Junker.
\newblock Learning factors analysis--a general method for cognitive model evaluation and improvement.
\newblock In \emph{International conference on intelligent tutoring systems}, pages 164--175. Springer, 2006.
\newblock \doi{10.1007/11774303_17}.

\bibitem[Corbalan et~al.(2008)Corbalan, Kester, and Van~Merri{\"e}nboer]{Corbalan2008Adaptation}
Gemma Corbalan, Liesbeth Kester, and Jeroen~JG Van~Merri{\"e}nboer.
\newblock Selecting learning tasks: Effects of adaptation and shared control on learning efficiency and task involvement.
\newblock \emph{Contemporary educational psychology}, 33\penalty0 (4):\penalty0 733--756, 2008.
\newblock \doi{10.1016/j.cedpsych.2008.02.003}.

\bibitem[Corbett and Anderson(1994)]{corbett1994bkt}
Albert~T Corbett and John~R Anderson.
\newblock Knowledge tracing: Modeling the acquisition of procedural knowledge.
\newblock \emph{User modeling and user-adapted interaction}, 4\penalty0 (4):\penalty0 253--278, 1994.
\newblock \doi{10.1007/BF01099821}.

\bibitem[Foster et~al.(2019)Foster, Mueller, Was, Rawson, and Dunlosky]{foster2019}
Nathaniel~L Foster, Michael~L Mueller, Christopher Was, Katherine~A Rawson, and John Dunlosky.
\newblock Why does interleaving improve math learning? the contributions of discriminative contrast and distributed practice.
\newblock \emph{Memory \& Cognition}, 47\penalty0 (6):\penalty0 1088--1101, 2019.
\newblock \doi{10.3758/s13421-019-00918-4}.

\bibitem[Holt(2024)]{holt2024five}
Laurence Holt.
\newblock The 5 percent problem.
\newblock \emph{Education Next}, 24\penalty0 (4):\penalty0 26--31, 2024.

\bibitem[Huang et~al.(2020)Huang, Aleven, McLaughlin, and Koedinger]{huang2020designloop}
Yun Huang, Vincent Aleven, Elizabeth McLaughlin, and Kenneth Koedinger.
\newblock A general multi-method approach to design-loop adaptivity in intelligent tutoring systems.
\newblock pages 124--129, 2020.
\newblock \doi{10.1007/978-3-030-52240-7_23}.

\bibitem[K{\"a}ser and Alexandron(2024)]{kaeser2024}
Tanja K{\"a}ser and Giora Alexandron.
\newblock Simulated learners in educational technology: A systematic literature review and a turing-like test.
\newblock \emph{International Journal of Artificial Intelligence in Education}, 34\penalty0 (2):\penalty0 545--585, 2024.
\newblock \doi{10.1007/s40593-023-00352-1}.

\bibitem[Koedinger et~al.(2015)Koedinger, D’Mello, Mclaughlin, Pardos, and Ros{\'e}]{Koedinger2015DataMining}
K.~Koedinger, Sidney~K. D’Mello, Elizabeth Mclaughlin, Zachary~A. Pardos, and Carolyn~Penstein Ros{\'e}.
\newblock Data mining and education.
\newblock \emph{Wiley interdisciplinary reviews. Cognitive science}, 6 4:\penalty0 333--353, 2015.
\newblock \doi{10.1002/wcs.1350}.

\bibitem[Koedinger et~al.(2010)Koedinger, Baker, Cunningham, Skogsholm, Leber, and Stamper]{koedinger2010data}
Kenneth~R Koedinger, Ryan~SJd Baker, Kyle Cunningham, Alida Skogsholm, Brett Leber, and John Stamper.
\newblock A data repository for the edm community: The pslc datashop.
\newblock \emph{Handbook of educational data mining}, 43:\penalty0 43--56, 2010.

\bibitem[Koedinger et~al.(2023)Koedinger, Carvalho, Liu, and McLaughlin]{Koedinger2023}
Kenneth~R Koedinger, Paulo~F Carvalho, Ran Liu, and Elizabeth~A McLaughlin.
\newblock An astonishing regularity in student learning rate.
\newblock \emph{Proceedings of the National Academy of Sciences}, 120\penalty0 (13):\penalty0 e2221311120, 2023.
\newblock \doi{10.1073/pnas.2221311120}.

\bibitem[Long and Aleven(2017)]{long2017}
Yanjin Long and Vincent Aleven.
\newblock Enhancing learning outcomes through self-regulated learning support with an open learner model.
\newblock \emph{User Modeling and User-Adapted Interaction}, 27:\penalty0 55--88, 2017.
\newblock \doi{10.1007/s11257-016-9186-6}.

\bibitem[MacLellan(2025)]{maclellan2025modelhumanlearnerscomputational}
Christopher~J MacLellan.
\newblock Model human learners: Computational models to guide instructional design.
\newblock \emph{arXiv preprint arXiv:2502.02456}, 2025.
\newblock \doi{10.48550/arXiv.2502.02456}.

\bibitem[Pel{\'a}nek(2017)]{Pelanek2017}
Radek Pel{\'a}nek.
\newblock Bayesian knowledge tracing, logistic models, and beyond: an overview of learner modeling techniques.
\newblock \emph{User modeling and user-adapted interaction}, 27\penalty0 (3):\penalty0 313--350, 2017.
\newblock \doi{10.1007/s11257-017-9193-2}.

\bibitem[Scheiter and Gerjets(2007)]{scheiter2007learner}
Katharina Scheiter and Peter Gerjets.
\newblock Learner control in hypermedia environments.
\newblock \emph{Educational Psychology Review}, 19\penalty0 (3):\penalty0 285--307, 2007.
\newblock \doi{10.1007/s10648-007-9046-3}.

\bibitem[Stamper et~al.(2013)Stamper, Eagle, Barnes, and Croy]{Stamper2013}
John Stamper, Michael Eagle, Tiffany Barnes, and Marvin Croy.
\newblock Experimental evaluation of automatic hint generation for a logic tutor.
\newblock \emph{International Journal of Artificial Intelligence in Education}, 22\penalty0 (1-2):\penalty0 3--17, 2013.
\newblock \doi{10.3233/JAI-130029}.

\bibitem[Urhahne and Wijnia(2023)]{urhahne2023}
Detlef Urhahne and Lisette Wijnia.
\newblock Theories of motivation in education: An integrative framework.
\newblock \emph{Educational Psychology Review}, 35\penalty0 (2):\penalty0 45, 2023.
\newblock \doi{10.1007/s10648-023-09767-9}.

\bibitem[Xia et~al.(2025)Xia, Schmucker, Borchers, and Aleven]{xia2025}
Meng Xia, Robin Schmucker, Conrad Borchers, and Vincent Aleven.
\newblock Optimizing mastery learning by fast-forwarding over-practice steps.
\newblock In \emph{European Conference on Technology Enhanced Learning}, pages 549--563. Springer, 2025.
\newblock \doi{10.1007/978-3-032-03870-8_37}.

\bibitem[Yan et~al.(2024)Yan, Lin, and Kinshuk]{yan2024control}
Hongxin Yan, Fuhua Lin, and Kinshuk.
\newblock An ai-learner shared control model design for adaptive practicing.
\newblock In \emph{International Conference on Intelligent Tutoring Systems}, pages 272--280. Springer, 2024.
\newblock \doi{10.1007/978-3-031-63028-6_21}.

\end{thebibliography}

\balancecolumns
\end{document}